\documentclass[12pt]{article}
\usepackage{amsmath, amssymb, graphics,geometry}
\newcommand{\mathsym}[1]{{}}

\title{Two standard methods for solving the Ito equation}

\author{Alvaro Salas\thanks{Department of Mathematics, Universidad de
Caldas, Department of Mathematics, Universidad Nacional de Colombia,
Manizales, Colombia. \emph{email} : asalash2002@yahoo.com}}
\date{}
\begin {document}
\maketitle
\begin {abstract}
In this paper we show some exact solutions for the Ito equation.
These solutions are obtained by two wel known methods: the tanh method and the
projective Riccati equation method.
\end{abstract}
\circledR Mathematica.
\section{Introduction}
A large variety of physical, chemical, and biological phenomena is
governed by nonlinear evolution equations. The analytical study of
nonlinear partial differential equations was of great interest
during the last decades. Investigations of traveling wave solutions
of nonlinear equations play an important role in the study of
nonlinear physical phenomena. The importance of obtaining the exact
solutions, if available, of those nonlinear equations facilitates
the verification of numerical solvers and aids in the stability
analysis of solutions. In this paper we show solutions for Ito's
equation by using the thanh method [3] and the projective Riccati
equation method [1]. This equation is a particular case of  the
general fifth order KdV equation (fKdV)
\begin{equation}\label{eq0}
u_t+\omega\, u_{xxxxx}+\alpha u^2u_x+\beta\, u_xu_{xx}\, +\gamma
\,uu_{xxx}=0,
\end{equation}
where $\alpha$, $\beta$, $\gamma$ and $\omega$ are arbitrary real
parameters. Ito's equation results from (\ref{eq0}) when $\alpha=2$,
$\beta=6$, $\gamma=3$ and $\omega=1$ . Thus, the Ito equation reads
\begin{equation}\label{eq1}
u_t+ u_{xxxxx}+2 u^2u_x+6 u_xu_{xx}\, +3 \,uu_{xxx}=0,
\end{equation}
To solve the equation (\ref{eq1}) we first unite the independent
variables $x$ and $t$ into one wave variable $\xi=x+\lambda\,t$ to
carry the PDE (\ref{eq1}) into the ODE (ordinary differential
equation)
\begin{equation}\label{eq1a}
2 v'(\xi ) v(\xi )^2+3
   v^{(3)}(\xi ) v(\xi
   )+\lambda  v'(\xi )+6
   v'(\xi ) v''(\xi
   )+v^{(5)}(\xi )=0
\end{equation}
so that $u(x,t)=v(\xi)=v(x+\lambda\,t)$.
\section{Solution by the tanh method}
We seek solutions of equation (\ref{eq1a}) in the form
\begin{equation}\label{eq1b}
v(\xi)=\sum_{j=0}^M\,a_j\varphi^j(\xi),
\end{equation}
where the function $\varphi(\xi)$ satisfies the Riccati equation
\begin{equation}\label{eq1c}
\varphi\,'(\xi)=k+\varphi^2(\xi)
\end{equation}

This equation admits the following solutions :
\begin{eqnarray}
\varphi(\xi)=\sqrt{k}\,\tan{\sqrt{k}\,\xi}&\text{and}&\varphi(\xi)=-\sqrt{k}\,\cot{\sqrt{k}\,\xi}\quad \text{for}\quad k>0.\label{eq1d}\\
\varphi(\xi)=-\sqrt{-k}\,\tanh{\sqrt{-k}\,\xi}&\text{and}&\varphi(\xi)=-\sqrt{-k}\,\coth{\sqrt{-k}\,\xi}\quad \text{for}\quad k>0.\label{eq1e}\\
&&\varphi(\xi)=-\dfrac{1}{\xi}\quad \text{for}\quad k=0.\label{eq1f}
\end{eqnarray}
Substituting (\ref{eq1b}) and (\ref{eq1c}) into the ODE (\ref{eq1a})
results in a polynomial equation in powers of $\varphi(\xi)$. To
determine the parameter $M$, we usually balance the linear terms of
highest order in the resulting equation with the highest order
nonlinear terms. The highest order of $v^{(p)}(\xi)$ is $M + p$. We
write this in the form
\begin{equation}\label{eq1f}
O(v^{(p)}(\xi))=O(v^{p})=M+p.\quad{\text{In particular,}}\quad
O(v(\xi))=O(v)=M+0=M.
\end{equation}
The following formulae are useful in the balancing process :
\begin{equation}\label{eq1g}
O(v^{(p)}v^{(q)})=2M+p+q;\quad O(v^pv^{(q)})=(p+1)M+q.
\end{equation}
With $M$ determined, we collect all coefficients of powers of
$\varphi(\xi)$ in the resulting equation where these coefficients
have to vanish. This will give a system of algebraic equations
involving the parameters $a_i$, $(i = 0,1,\ldots,M)$, $\lambda$, and
$k$. Having determined these parameters, knowing that $M$ is a
positive integer in most cases, and using (\ref{eq1b}) we obtain an
analytic solution. From (\ref{eq1f}) and (\ref{eq1g}) it is clear
that $O(v^2v')=O(v^{2}v^{(1)})=3M+1$ and $O(v^{(5)})=M+5$. Balancing
$v^2(\xi)v'(\xi)=v^2v'$ with $v^{(5)}(\xi)=v^{(5)}$ in (\ref{eq1a})
we find $3M+1=M+5$, that is, $M=2$. Thus, we look for solutions of
(\ref{eq1a}) in the form
\begin{equation}\label{eq1h}
v(\xi)=a_0+a_1\varphi(\xi)+a_2\varphi^2(\xi).
\end{equation}
After substitution of the equations (\ref{eq1h}) and (\ref{eq1c}) in
(\ref{eq1a}), we obtain the following system of nonlinear algebraic
equations :
$$
\begin{array}{ll}
 (1) & 16 a_1 k^3+12 a_1 a_2
   k^3+6 a_0 a_1 k^2+2 a_0^2
   a_1 k+\lambda  a_1 k=0 \\
 (2) & 2 k a_1^3+136 k^2 a_1+2
   a_0^2 a_1+\lambda  a_1+24 k
   a_0 a_1+138 k^2 a_2 a_1+12
   k a_0 a_2 a_1=0 \\
 (3) & 24 a_2^2 k^3+272 a_2
   k^3+18 a_1^2 k^2+48 a_0 a_2
   k^2+4 a_0 a_1^2 k+4 a_0^2
   a_2 k+2 \lambda  a_2 k=0 \\
  (4) & 168 a_2^2 k^2+1232 a_2
   k^2+48 a_1^2 k+8 a_0 a_2^2
   k+8 a_1^2 a_2 k+120 a_0 a_2
   k+4 a_0 a_1^2+\\
   &4 a_0^2 a_2+2
   \lambda  a_2=0 \\
  (5) & 10 a_1 a_2^2+150 a_1
   a_2+120 a_1=0 \\
  (6) & 2 a_1^3+10 k a_2^2
   a_1+240 k a_1+18 a_0
   a_1+276 k a_2 a_1+12 a_0
   a_2 a_1=0 \\
 (7) & 4 a_2^3+144 a_2^2+720
   a_2=0 \\
 (8) & 4 k a_2^3+288 k a_2^2+8
   a_0 a_2^2+8 a_1^2 a_2+1680
   k a_2+72 a_0 a_2+30 a_1^2=0
\end{array}
$$
After solving this system, we get the following solutions
:\newline\\
\textbf{a.} $a_0=-5\sqrt{-\dfrac{\lambda}{6}}$, $a_1=0$, $a_2=-30$,
$k=\dfrac{1}{4}\sqrt{-\dfrac{\lambda}{6}}$ and
\begin{equation}\label{eq1i}
u_1(x,t)=
  - \dfrac{5}{2}\sqrt{-\dfrac{\lambda}{6}}\left(
      2 + 3\, \text{tan}^2\left(
          \dfrac{1}{2}\sqrt[4]{-\dfrac{\lambda}{6}}\, \xi
            \right)
      \right).
\end{equation}
\textbf{b.} $a_0=-5\sqrt{-\dfrac{\lambda}{6}}$, $a_1=0$, $a_2=-30$,
$k=\dfrac{1}{4}\sqrt{-\dfrac{\lambda}{6}}$ and
\begin{equation}\label{eq1j}
u_2(x,t)=
  - \dfrac{5}{2}\sqrt{-\dfrac{\lambda}{6}}\left(
      2 + 3\, \text{cot}^2\left(
          \dfrac{1}{2}\sqrt[4]{-\dfrac{\lambda}{6}}\, \xi
            \right)
      \right).
\end{equation}
\textbf{c.} $a_0=5\sqrt{-\dfrac{\lambda}{6}}$, $a_1=0$, $a_2=-30$,
$k=-\dfrac{1}{4}\sqrt{-\dfrac{\lambda}{6}}$ and
\begin{equation}\label{eq1j}
u_3(x,t)=
  \dfrac{5}{2}\sqrt{-\dfrac{\lambda}{6}}\left(
      2 - 3\, \text{tanh}^2\left(
          \dfrac{1}{2}\sqrt[4]{-\dfrac{\lambda}{6}}\, \xi
            \right)
      \right).
\end{equation}
\textbf{d.} $a_0=5\sqrt{-\dfrac{\lambda}{6}}$, $a_1=0$, $a_2=-30$,
$k=-\dfrac{1}{4}\sqrt{-\dfrac{\lambda}{6}}$ and
\begin{equation}\label{eq1j}
u_4(x,t)=
  \dfrac{5}{2}\sqrt{-\dfrac{\lambda}{6}}\left(
      2 - 3\, \text{coth}^2\left(
          \dfrac{1}{2}\sqrt[4]{-\dfrac{\lambda}{6}}\, \xi
            \right)
      \right).
\end{equation}

\section{Solutions by the projective Riccati equation method}
This method is due to Conte[1]. \\
\textit{\textbf{Step 1}}. We consider solutions of (\ref{eq1a}) in
the form
\begin{equation}\label{eq4}
v(\xi)=a_0+\sum_{j=1}^m
\sigma(\xi)^{j-1}(a_j\sigma(\xi)+b_j\tau(\xi)),
\end{equation}
where $\sigma(\xi)$, $\tau(\xi)$  satisfy the system
\begin{equation}\label{eq5}
\begin{cases}
\sigma'(\xi)=e\sigma(\xi)\tau(\xi)\\
\tau'(\xi)=e\tau^2(\xi)-\mu \sigma(\xi)+r.
\end{cases}
\end{equation}
It may be proved that the first integral of this system is given by
\begin{equation}\label{eq6}
\tau^2=-e\left[r-2 \mu
\sigma(\xi)+\dfrac{\mu^2+\rho}{r}\sigma^2(\xi)\right],
\end{equation}
where $\rho=\pm 1$ and $e=\pm 1$. \\\noindent We consider the
following solutions of the system (\ref{eq5}).
\begin{enumerate}
\item \textbf{Case I}:
\newline
If $r=\mu =0$ then
\begin{equation}
\tau_1(\xi)=-\frac{1}{e\xi},\;\;\sigma_1(\xi)=\frac{C}{\xi}.
\end{equation}
\item \textbf{Case II}:
\newline
If $e=1$, $\rho=-1$ and $r>0$ :
\begin{equation}
\begin{cases}
\sigma_1(\xi )=
    \dfrac{r\,\sec ({\sqrt{r}}\,\xi )}
  {1 + \mu \,\sec ({\sqrt{r}}\,\xi )};\quad
\tau_1(\xi )=
   \dfrac{{\sqrt{r}}\,\tan ({\sqrt{r}}\,\xi )}
  {1 + \mu \,\sec ({\sqrt{r}}\,\xi )}\\\\
\sigma_2(\xi )=\dfrac{r\,\csc ({\sqrt{r}}\,\xi )}
  {1 + \mu \,\csc ({\sqrt{r}}\,\xi )};\quad
\tau_2(\xi )=-\dfrac{{\sqrt{r}}\,\cot ({\sqrt{r}}\,\xi )}
  {1 + \mu \,\csc ({\sqrt{r}}\,\xi )}.
\end{cases}
\end{equation}
\item \textbf{Case III}:
\newline
If $e=-1$, $\rho=-1$ and $r>0$ :
\begin{equation}
\begin{cases}
\sigma_3= \dfrac{r\,\text{sech}({\sqrt{r}}\,\xi )}
  {1 + \mu \,\text{sech}({\sqrt{r}}\,\xi )}\\\\
\tau_3=\dfrac{{\sqrt{r}}\,\tanh ({\sqrt{r}}\,\xi )}
  {1 + \mu \,\text{sech}({\sqrt{r}}\,\xi )}.
\end{cases}
\end{equation}
\item \textbf{Case IV}:
\newline
If $e=-1$, $\rho=1$ and $r>0$ :
\begin{equation}
\begin{cases}
\sigma_4= \dfrac{r\,\text{csch}({\sqrt{r}}\,\xi )}
  {1 + \mu \,\text{csch}({\sqrt{r}}\,\xi )}\\\\
\tau_4=\dfrac{{\sqrt{r}}\,\coth ({\sqrt{r}}\,\xi )}
  {1 + \mu \,\text{csch}({\sqrt{r}}\,\xi )}.
\end{cases}
\end{equation}
\end{enumerate}
$$$$
\noindent\textit{\textbf{Step 2}}. Substituting (\ref{eq4}), along
with (\ref{eq5}) and (\ref{eq6}) into (\ref{eq1a}) and collecting
all terms with the same power in $\sigma^{i}(\xi)\tau^{j}(\xi)$,  we
get a polynomial in the two variables $\sigma(\xi)$ and $\tau(\xi)$.
This polynomial has the form
\begin{eqnarray}
a\sigma(\xi)^{m+5}+b\sigma(\xi)^{2m+3}+c\sigma(\xi)^{m+4}\tau(\xi)+d\sigma(\xi)^{3m+1}+\nonumber\\
+e\sigma(\xi)^{2m+2}\tau(\xi)+\text{other terms of lower
degree}\label{eq7}
\end{eqnarray}
We assume that $m\geq 1$ to avoid trivial solutions.  The degrees of
the highest terms are $m+5$ ( the degree of the terms
$a\sigma(\xi)^{m+5}$ and $c\sigma(\xi)^{m+4}\tau(\xi)$), $2m+3$ (
the degree of the term $b\sigma(\xi)^{2m+3}$ ) and $3m+1$ ( the
degree of the term $d\sigma(\xi)^{3m+1}$ ). There are two integer
values of $m$ for which $3m+1=2m+3$ or $3m+1=m+5$ or $2m+3=m+5$.
These are $m=1$ and $m=2$. We are going to find solutions for $m=1$
( the case $m=2$ is not considered here ).  When $m=1$  solutions
have the form $v(\xi)=a_0+ a_1\sigma(\xi)+b_1\tau(\xi)$ and equating
in (\ref{eq7}) the coefficients of every power of $\sigma(\xi)$ and
of every term of the form $\sigma^j(\xi)\tau(\xi)$ to zero, we
obtain the following algebraic system in the variables $a_0$, $a_1$,
$b_1,\ldots$ : \newline\newline\noindent
$\begin{array}{ll}
 (1) & e^7 \left(\mu ^2+\rho
   \right)^2 a_1=0 \\
 \end{array}$\newline
 $\begin{array}{ll}
 (2) & e^8 \left(\mu ^2+\rho
   \right)^3 b_1=0 \\
 \end{array}$\newline
 $\begin{array}{ll}
 (3) & e^5 \left(\mu ^2+\rho
   \right)^2 \left(12 \mu
   e^3-6 \mu  e+a_1\right)
   b_1=0 \\
 \end{array}$\newline
 $\begin{array}{ll}
(4) & e^4 \left(\mu ^2+\rho
   \right) \left(16 r \mu  a_1
   e^3-\mu ^2 b_1^2 e-\rho
   b_1^2 e-8 r \mu  a_1 e+r
   a_1^2\right)=0 \\
 \end{array}$\newline
 $\begin{array}{ll}
(5) & 2 (e-1) (e+1) r \left(15
   r e^3-9 r e-2 a_0\right)
   b_1^2=0\\
 \end{array}$\newline
 $\begin{array}{ll}
 (6) & 120 r^2 a_1 e^7-120 r \mu
    b_1^2 e^5-180 r^2 a_1
   e^5-18 r a_0 a_1 e^4+162 r
   \mu  b_1^2 e^3+61 r^2 a_1
   e^3+8 \mu  a_0 b_1^2 e^2\\
   &-6
   r a_1 b_1^2 e^2+15 r a_0
   a_1 e^2-45 r \mu  b_1^2 e+2
   a_0^2 a_1 e+\lambda  a_1
   e-4 \mu  a_0 b_1^2+4 r a_1
   b_1^2=0 \\
 \end{array}$\newline
 $\begin{array}{ll}
 (7) & -480 r^2 \mu  a_1 e^7+180
   r \mu ^2 b_1^2 e^5+60 r
   \rho  b_1^2 e^5+600 r^2 \mu
    a_1 e^5-30 r^2 a_1^2
   e^4+36 r \mu  a_0 a_1
   e^4-180 r \mu ^2 b_1^2
   e^3\\
   &-48 r \rho  b_1^2
   e^3-150 r^2 \mu  a_1 e^3+21
   r^2 a_1^2 e^2-4 \mu ^2 a_0
   b_1^2 e^2-4 \rho  a_0 b_1^2
   e^2+12 r \mu  a_1 b_1^2
   e^2\\
   &-18 r \mu  a_0 a_1 e^2+4
   r a_0 a_1^2 e+27 r \mu ^2
   b_1^2 e-4 r \mu  a_1
   b_1^2=0 \\
 \end{array}$\newline
 $\begin{array}{ll}
 (8) & e (\,360 r \mu ^2 a_1
   e^6+120 r \rho  a_1 e^6-60
   \mu ^3 b_1^2 e^4-60 \mu
   \rho  b_1^2 e^4-330 r \mu
   ^2 a_1 e^4-90 r \rho  a_1
   e^4+30 r \mu  a_1^2 e^3\\
   &-9
   \mu ^2 a_0 a_1 e^3-9 \rho
   a_0 a_1 e^3+33 \mu ^3 b_1^2
   e^2+33 \mu  \rho  b_1^2
   e^2+45 r \mu ^2 a_1 e^2-12
   r \mu  a_1^2 e-3 \mu ^2 a_1
   b_1^2 e\\
   &-3 \rho  a_1 b_1^2
   e+r a_1^3\,)=0 \\
 \end{array}$\newline
 $\begin{array}{ll}
 (9) & (e-1) (e+1) r b_1
   \left(120 r^2 e^6-120 r^2
   e^4-18 r a_0 e^3+16 r^2
   e^2-2 r b_1^2 e+6 r a_0 e+2
   a_0^2+\lambda \right)=0 \\
 \end{array}$\newline
 $\begin{array}{ll}
 (10) & b_1 (\,720 r^2 \mu
   e^8-1320 r^2 \mu  e^6-72 r
   \mu  a_0 e^5+60 r^2 a_1
   e^5+662 r^2 \mu  e^4-8 r
   \mu  b_1^2 e^3+84 r \mu
   a_0 e^3\\
   &-69 r^2 a_1 e^3+4
   \mu  a_0^2 e^2-61 r^2 \mu
   e^2+2 \lambda  \mu  e^2-8 r
   a_0 a_1 e^2+6 r \mu  b_1^2
   e-15 r \mu  a_0 e+12 r^2
   a_1 e\\
   &-2 \mu  a_0^2-\lambda
   \mu +4 r a_0 a_1\,)=0
  \\
 \end{array}$\newline
 $\begin{array}{ll}
(11) & b_1 (\,1800 r^2 \mu
   ^2 e^8+360 r^2 \rho
   e^8-2880 r^2 \mu ^2 e^6-480
   r^2 \rho  e^6-108 r \mu ^2
   a_0 e^5-36 r \rho  a_0
   e^5+\\
   &240 r^2 \mu  a_1
   e^5+1186 r^2 \mu ^2 e^4+136
   r^2 \rho  e^4-12 r \mu ^2
   b_1^2 e^3-4 r \rho  b_1^2
   e^3+96 r \mu ^2 a_0 e^3+24
   r \rho  a_0 e^3\\
   &-228 r^2 \mu
    a_1 e^3-75 r^2 \mu ^2
   e^2+\lambda  \mu ^2 e^2+2
   \mu ^2 a_0^2 e^2+2 \rho
   a_0^2 e^2+6 r^2 a_1^2
   e^2+\lambda  \rho  e^2-16 r
   \mu  a_0 a_1 e^2\\
   &+6 r \mu ^2
   b_1^2 e+2 r \rho  b_1^2 e-9
   r \mu ^2 a_0 e+27 r^2 \mu
   a_1 e-2 r^2 a_1^2+4 r \mu
   a_0 a_1\,)=0 \\
 \end{array}$\newline
 $\begin{array}{ll}
 (12) & b_1 (\,2400 r \mu ^3
   e^8+1440 r \mu  \rho
   e^8-3120 r \mu ^3 e^6-1680
   r \mu  \rho  e^6-72 \mu ^3
   a_0 e^5-72 \mu  \rho  a_0
   e^5\\
   &+360 r \mu ^2 a_1
   e^5+120 r \rho  a_1 e^5+930
   r \mu ^3 e^4+390 r \mu
   \rho  e^4-8 \mu ^3 b_1^2
   e^3-8 \mu  \rho  b_1^2
   e^3+36 \mu ^3 a_0 e^3\\
   &+36
   \mu  \rho  a_0 e^3-249 r
   \mu ^2 a_1 e^3-69 r \rho
   a_1 e^3-30 r \mu ^3 e^2+12
   r \mu  a_1^2 e^2-8 \mu ^2
   a_0 a_1 e^2-8 \rho  a_0 a_1
   e^2\\
   &+2 \mu ^3 b_1^2 e+2 \mu
   \rho  b_1^2 e+15 r \mu ^2
   a_1 e-2 r \mu
   a_1^2\,)=0 \\
 \end{array}$\newline
 $\begin{array}{ll}
 (13) & 2 e^2 (\,\mu ^2+\rho
   \,) b_1 (\,900 r \mu
   ^2 e^6+180 r \rho  e^6-840
   r \mu ^2 e^4-120 r \rho
   e^4-9 \mu ^2 a_0 e^3-9 \rho
    a_0 e^3+120 r \mu  a_1
   e^3\\
   &+135 r \mu ^2 e^2-\mu ^2
   b_1^2 e-\rho  b_1^2 e-45 r
   \mu  a_1 e+3 r
   a_1^2\,)=0
\end{array}$

$$\,$$
\noindent\textit{\textbf{Step 3}}. ( This is the more difficult step
) Solving the previous system for $r$,
 $\mu$, $a_0$, $a_1$, $b_1$ with the aid of \circledR\emph{Mathematica 6.0} we get $b_1=0$, so solutions have
 the form $u=a_0+a_1\sigma(x+\lambda\,t)$. These solutions
 correspond to values $e=1$ and $\rho=-1$. They are :$$\,$$
\textbf{i.} $a_0=\dfrac{5\sqrt{-\lambda}}{2\sqrt{6}}$, $a_1=15$,
$\mu=-1$, $r=\sqrt{-\dfrac{\lambda}{6}}$ and
\begin{equation}\label{eq7}
u_5(x,t)=\dfrac{5\sqrt{-\lambda}}{2\sqrt{6}}\cdot\dfrac{1+5\,\text{csc}\left(\sqrt[4]{-\dfrac{\lambda}{6}}\,(x+\lambda\,t)\right)}{1-\,\text{csc}\left(\sqrt[4]{-\dfrac{\lambda}{6}}\,(x+\lambda\,t)\right)}
\end{equation}
\textbf{ii.} $a_0=\dfrac{5\sqrt{-\lambda}}{2\sqrt{6}}$, $a_1=-15$,
$\mu=1$, $r=\sqrt{-\dfrac{\lambda}{6}}$ and
\begin{equation}\label{eq8}
u_6(x,t)=\dfrac{5\sqrt{-\lambda}}{2\sqrt{6}}\cdot\dfrac{1-5\,\text{csc}\left(\sqrt[4]{-\dfrac{\lambda}{6}}\,(x+\lambda\,t)\right)}{1+\,\text{csc}\left(\sqrt[4]{-\dfrac{\lambda}{6}}\,(x+\lambda\,t)\right)}
\end{equation}
\textbf{iii.} $a_0=\dfrac{5\sqrt{-\lambda}}{2\sqrt{6}}$, $a_1=-15$,
$\mu=1$, $r=\sqrt{-\dfrac{\lambda}{6}}$ and
\begin{equation}\label{eq9}
u_7(x,t)=\dfrac{5\sqrt{-\lambda}}{2\sqrt{6}}\left(1-3\,\text{sec}^2\left(\dfrac{1}{2}\sqrt[4]{-\dfrac{\lambda}{6}}\,(x+\lambda\,t)\right)\right)
\end{equation}
\textbf{iv.} $a_0=\dfrac{5\sqrt{-\lambda}}{2\sqrt{6}}$, $a_1=15$,
$\mu=-1$, $r=\sqrt{-\dfrac{\lambda}{6}}$ and
\begin{equation}\label{eq10}
u_8(x,t)=\dfrac{5\sqrt{-\lambda}}{2\sqrt{6}}\left(1-3\,\text{csc}^2\left(\dfrac{1}{2}\sqrt[4]{-\dfrac{\lambda}{6}}\,(x+\lambda\,t)\right)\right)
\end{equation}
\textbf{V.} $a_0=-\dfrac{5\sqrt{-\lambda}}{2\sqrt{6}}$, $a_1=-15$,
$\mu=1$, $r=-\sqrt{-\dfrac{\lambda}{6}}$ and
\begin{equation}\label{eq11}
u_9(x,t)=-\dfrac{5\sqrt{-\lambda}}{2\sqrt{6}}\left(1-3\,\text{sech}^2\left(\dfrac{1}{2}\sqrt[4]{-\dfrac{\lambda}{6}}\,(x+\lambda\,t)\right)\right)
\end{equation}
\textbf{Vi.} $a_0=-\dfrac{5\sqrt{-\lambda}}{2\sqrt{6}}$, $a_1=15$,
$\mu=-1$, $r=-\sqrt{-\dfrac{\lambda}{6}}$ and
\begin{equation}\label{eq12}
u_{10}(x,t)=-\dfrac{5\sqrt{-\lambda}}{2\sqrt{6}}\left(1+3\,\text{csch}^2\left(\dfrac{1}{2}\sqrt[4]{-\dfrac{\lambda}{6}}\,(x+\lambda\,t)\right)\right)
\end{equation}

\section{Conclusions}
In this paper, by using the tanh method, the projective Riccati
equation method and the help of symbolic computation system
\emph{Mathematica 6.0}, we obtained some exact solutions for the
equation (\ref{eq1}).  The projective Riccati equation method is
more complicated than other methods in comparation with the tanh
method  in the sense that the algebraic system in many cases demands
a lot of time to be solved. On the other hand, this method allows us
to obtain some new exact solutions. In fact, the projective method
gave us six solutions, while with the tanh method we obtained four
solutions. On the other hand, the solutions obtained by these
methods are different. We apply the mentioned method to solve other
nonlinear differential equations [2].

\end{document}